\documentclass[
amsmath,
prd,nofootinbib,floatfix,12pt,
]{revtex4}

\def\MSbar{\overline{\rm MS}}

\newcommand\beq{\begin{eqnarray}}
\newcommand\eeq{\end{eqnarray}}
\newcommand\Tbar{\overline{T}}
\def\lsim{\mathrel{\rlap{\lower4pt\hbox{$\sim$}}
    \raise1pt\hbox{$<$}}}                
\def\gsim{\mathrel{\rlap{\lower4pt\hbox{$\sim$}}
    \raise1pt\hbox{$>$}}}            

\allowdisplaybreaks
\usepackage{graphicx}
\usepackage{setspace}
\usepackage{bm}

\begin{document}

\renewcommand{\theequation}{\arabic{section}.\arabic{equation}}
\renewcommand{\thefigure}{\arabic{section}.\arabic{figure}}
\renewcommand{\thetable}{\arabic{section}.\arabic{table}}

\title{\large
$Z$ boson pole mass at two-loop order in the pure $\overline{\rm MS}$ scheme
}\baselineskip=16pt 

\author{Stephen P.~Martin}
\affiliation{
{\it Department of Physics, Northern Illinois University, DeKalb IL 60115},
{\it Fermi National Accelerator Laboratory, P.O. Box 500, Batavia IL 60510}}

\begin{abstract}\normalsize\baselineskip=16pt 
I obtain the complex pole squared mass of the $Z$ boson 
at full two-loop order in the Standard Model in the pure $\MSbar$ 
renormalization scheme. The input parameters are the 
running gauge couplings, the top-quark Yukawa coupling, the Higgs self-coupling,
and the vacuum expectation value that minimizes the Landau gauge
effective potential. The effects of non-zero
Goldstone boson mass are resummed. Within a reasonable range of renormalization 
scale choices, the scale dependence of the computed 
pole mass is found to be comparable to the current experimental uncertainty, but the 
true theoretical error is likely somewhat larger. 
\end{abstract}
\maketitle

\vspace{-0.45in}

\tableofcontents

\baselineskip=16pt

\vspace{-0.15in}

\section{Introduction\label{sec:intro}}
\setcounter{equation}{0}
\setcounter{figure}{0}
\setcounter{table}{0}
\setcounter{footnote}{1}

One of the cornerstone physical observables of the Standard Model is the $Z$ boson mass.
The experimental value that is usually quoted is obtained using a Breit-Wigner 
parametrization with a variable width,  and is given in 
ref.~\cite{RPP} from a fit to LEP data as:
\beq
M_Z^{\rm exp} = 91.1876 \pm 0.0021\>{\rm GeV}.
\eeq
This is related \cite{Bardin:1988xt,Willenbrock:1991hu,Sirlin:1991fd} 
to the real part of the 
complex pole squared mass $s^Z_{\rm pole} = M_Z^2 - i \Gamma_Z M_Z$
(with $\Gamma_Z$ a constant width) according 
to:
\beq
M_Z &=& M_Z^{\rm exp} (1 - \Gamma_Z^2/2 M_Z^2 + \ldots) 
\label{eq:poleBW}
\\
&=& 91.1535 \pm 0.0021\>{\rm GeV} .
\label{eq:MZpoleLEP}
\eeq
In general, the complex pole squared mass is a physical 
observable 
\cite{Tarrach:1980up,Stuart:1991xk,Stuart:1992jf,Passera:1996nk,
Kronfeld:1998di,Gambino:1999ai}, 
independent of the choice of renormalization scheme and scale 
and the choice of gauge fixing.

In this paper, I report a calculation, at full 2-loop order, 
of the complex pole squared 
mass parameters $M_Z$ and $\Gamma_Z$, 
using the pure $\MSbar$ scheme. The input parameters 
in this scheme are the running renormalized quantities
\beq
g, g', g_3,  y_t, \lambda, v,
\eeq
where the first three are the Standard Model gauge couplings, $y_t$ is the top-quark Yukawa coupling, $\lambda$ is the Higgs self-coupling, and 
the vacuum expectation value (VEV) $v$ is defined here to be the 
minimum of the full radiatively corrected
effective potential in the Landau gauge.
The normalizations used here for $\lambda$ and $v$ are fixed by
writing the tree-level Higgs potential as
\beq
V(\Phi,\Phi^\dagger) &=& m^2 \Phi^\dagger \Phi + \lambda (\Phi^\dagger \Phi)^2 ,
\eeq
where the canonically normalized doublet Higgs field has VEV
$\langle \Phi \rangle = v/\sqrt{2}$, and $m^2$ is a negative Higgs squared mass parameter. 
The minimization condition that relates $v$ to $m^2$ 
(allowing the latter to be eliminated)
is presently known at full 2-loop order 
\cite{FJJ} augmented by all 3-loop 
contributions at leading orders in both $g_3$ and $y_t$ \cite{Martin:2013gka}. 
Goldstone boson mass effects are resummed in this relation using
\cite{Martin:2014bca,Elias-Miro:2014pca}; this effect is
usually numerically small but is conceptually important, 
and in any case leads to simpler formulas. 

Other definitions of the Higgs VEV can be found in the literature.
One alternative (for example, see 
refs.~\cite{Jegerlehner:2001fb,Jegerlehner:2002em,Kniehl:2015nwa}) is to instead define the VEV as the minimum of the tree-level potential,
so $v_{\rm tree} = \sqrt{-m^2/\lambda}$. This has the disadvantage 
that one must include tadpole diagrams explicitly. Also, one is then
expanding around a point that differs from the true radiatively corrected vacuum, so 
perturbation theory converges less quickly, at least formally and for generic
choices of the renormalization scale. Indeed, in the
large $y_t$ limit, the loop expansion parameter is $N_c y_t^4/(16 \pi^2 \lambda)$ rather
than the usual $N_c y_t^2/16 \pi^2$. [For more details, see for example 
refs.~\cite{Martin:2014bca,Martin:2015lxa}, and the discussion surrounding
eq.~(\ref{eq:compJKV}) below.]
The reason for the $\lambda$ in the denominator is that the 
tadpole diagrams have a
Higgs propagator at zero momentum, which is just the reciprocal of the
Higgs squared mass.

Another alternative (see for example ref.~\cite{Degrassi:2014sxa}) 
is to define the VEV so that the sum of tadpole graphs in Feynman gauge vanishes. 
However, the Landau gauge effective potential is easier to compute 
to higher orders, and avoids renormalization of the gauge-fixing parameter, 
making it arguably a more convenient choice as a standard.

The pure $\MSbar$ scheme is an alternative to on-shell and hybrid schemes, 
which have been used for many precision studies of the $Z$ mass and the
electroweak sector.
For a selection of some important related results in that approach, see 
refs.~\cite{Sirlin:1980nh, Marciano:1980pb, Sirlin:1983ys, Djouadi:1987gn,
Consoli:1989fg, Kniehl:1989yc, Djouadi:1993ss, Avdeev:1994db,
Chetyrkin:1995ix, Chetyrkin:1995js, Degrassi:1996mg, Degrassi:1996ps,
Degrassi:1997iy, Passera:1998uj, Freitas:2000gg,
Awramik:2002wn, Faisst:2003px, Awramik:2003ee, Awramik:2003rn, Schroder:2005db,
Chetyrkin:2006bj, Boughezal:2006xk}, and for 
reviews see refs.~\cite{RPP,Sirlin:2012mh}.

\section{Complex pole mass of the $Z$ boson at 2-loop order\label{sec:MZ}}
\setcounter{equation}{0}
\setcounter{figure}{0}
\setcounter{table}{0}
\setcounter{footnote}{1}

To obtain the $Z$ boson pole squared mass, one begins with the symmetric 
$2 \times 2$ matrix of neutral gauge boson transverse self-energy functions, 
for $V,V' = \gamma,Z$: 
\beq
\Pi_{VV'}(s) &=& \frac{1}{16 \pi^2} \Pi^{(1)}_{VV'}(s) \,+\, 
\frac{1}{(16 \pi^2)^2} \Pi^{(2)}_{VV'}(s) \,+\, \ldots
\eeq
where $s = -p^2$, with $p^\mu$ the external momentum, using a metric with
Euclidean or ($-$,$+$,$+$,$+$) signature. 
These are obtained by calculating, in the theory in $d = 4 - 2\epsilon$ 
dimensions with bare parameters,
the sum of 1-particle-irreducible 2-point Feynman diagrams for $\Pi_{VV'}^{\mu\nu}$, followed by projecting with 
$(\eta_{\mu\nu} - p_\mu p_\nu/p^2)/(d-1)$.
The pole squared mass is then the solution of
\beq
s^Z_{\rm pole} &=& Z_B + \Pi_{ZZ}(s^Z_{\rm pole}) +
[\Pi_{\gamma Z}(s^Z_{\rm pole})]^2/\left [s^Z_{\rm pole} 
- \Pi_{\gamma\gamma}(s^Z_{\rm pole}) \right ].
\label{eq:polebare}
\eeq
Here,
\beq
Z_B = (g_B^2 + g_B^{\prime 2}) v_B^2/4
\eeq
is the bare, tree-level, squared mass of the $Z$ boson.
Solving eq.~(\ref{eq:polebare}) iteratively, one obtains to 2-loop order:
\beq
s^Z_{\rm pole} &=& Z_B \,+\, \frac{1}{16 \pi^2} \Pi^{(1)}_{ZZ}(Z_B) \,+\, 
\frac{1}{(16 \pi^2)^2} \Bigl \{ \Pi^{(2)}_{ZZ}(Z_B) \,+\, 
\Pi^{(1)\prime}_{ZZ}(Z_B)\, \Pi^{(1)}_{ZZ}(Z_B)
\nonumber \\ && + [\Pi^{(1)}_{\gamma Z}(Z_B)]^2/Z_B \Bigr \} .
\eeq
Instead of computing separate counterterm diagrams, the calculation 
described here was done in terms of only bare quantities $g_B$, $g'_B$, 
$g_{3B}$, $y_{tB}$, $\lambda_B$, $v_B$, $m^2_B$, and then translated to 
renormalized running $\MSbar$ quantities $g$, $g'$, $g_3$, $y_t$, $\lambda$, 
$v$ at the end. Tadpole diagrams need not be calculated, because 
they automatically sum to zero, due to the defining condition that the 
VEV is the minimum of the effective potential. Using the minimization 
condition for the Landau gauge effective potential given in 
ref.~\cite{Martin:2014bca}, the parameter $m^2$ (and the Goldstone boson 
squared mass) are eliminated. These procedures are the same as 
described in refs.~\cite{Martin:2014cxa,Martin:2015lxa}, and so most 
details will not be repeated here. An exception is that the 2-loop 
translation of the $U(1)_Y$ gauge couplings from bare to renormalized 
couplings is needed, to go along with eqs.~(2.5)-(2.24) of 
ref.~\cite{Martin:2014cxa} and eqs.~(2.3)-(2.10) of 
ref.~\cite{Martin:2015lxa}:
\beq
g_B' &=& \mu^\epsilon \Bigl [
g' + \frac{1}{16 \pi^2} \frac{c_{1,1}^{g'}}{\epsilon}
+ 
\frac{1}{(16\pi^2)^2} \Bigl ( \frac{c^{g'}_{2,2}}{\epsilon^2}
+ \frac{c^{g'}_{2,1}}{\epsilon} \Bigr ) 
+ \ldots \Bigr ]
\eeq
where
\beq
c^{g'}_{1,1} &=&  \frac{41}{12} g^{\prime 3}, 
\\
c^{g'}_{2,1} &=&  g^{\prime 3} \Bigl ( \frac{11}{3} g_3^2 + \frac{9}{8} g^2
+ \frac{199}{72} g^{\prime 2} - \frac{17}{24} y_t^2 \Bigr ),
\\
c^{g'}_{2,2} &=&  \frac{1681}{96} g^{\prime 5} ,
\eeq
and $\mu$ is the regularization scale, related to the renormalization scale $Q$ by
$
\mu^2 =  Q^2e^{\gamma_E}/4 \pi.
$
As in refs.~\cite{Martin:2014cxa,Martin:2015lxa},
the results are reduced, using the Tarasov recurrence relations \cite{Tarasov:1997kx} to 
a set of 1-loop basis integrals $A,B$ and 2-loop basis integrals 
$I,S,T,\Tbar,U,M$, following the notations and conventions 
of refs.~\cite{Martin:2003qz,TSIL}. The program TSIL \cite{TSIL} can be used to automatically and efficiently evaluate these basis integrals numerically. 
Where possible,  TSIL takes advantage of
analytical results in terms of polylogarithms, which were given 
in refs.~\cite{Martin:2003qz,TSIL,
Broadhurst:1987ei, Gray:1990yh, Davydychev:1992mt, Davydychev:1993pg, 
Scharf:1993ds, Berends:1994ed, Berends:1997vk}. 
In many cases, analytical results for the basis integrals 
are not available, so TSIL employs 
Runge-Kutta solution of differential equations in the external momentum invariant 
\cite{Martin:2003qz}, similar to that suggested in ref.~\cite{Caffo:1998du}. 

The final result for the 2-loop $Z$ boson complex pole mass can be written as: 
\beq
s^Z_{\rm pole}  &=& M^2_Z - i \Gamma_Z M_Z \>=\> 
Z
+ \frac{1}{16 \pi^2} \Delta^{(1)}_Z
+ \frac{1}{(16 \pi^2)^2} \left [
\Delta^{(2),{\rm QCD}}_Z + \Delta^{(2),{\rm non-QCD}}_Z \right ].
\phantom{xxx}
\label{eq:M2Zpole}
\eeq
In the following,
\beq
Z &=& (g^2 + g^{\prime 2}) v^2/4,
\\
W &=& g^2 v^2/4,
\\
t &=& y_t^2 v^2/2,
\\
h &=& 2 \lambda v^2
\eeq
are the tree-level $\MSbar$ squared masses of the $Z$ boson, $W$ boson, top quark, and Higgs boson, respectively, and the couplings
of the quarks and leptons to the $Z$ boson are: 
\beq
a_{f} &=& \sqrt{g^2 + g^{\prime 2}} \left [ T_3^f - Q_f \,
g^{\prime 2}/(g^2 + g^{\prime 2}) \right ],
\eeq
for $f = u_L, u_R, d_L, d_R, e_L, e_R, \nu_L$, where
\beq
&&
T_3^{u_L} = T_3^{\nu_L} = -T_3^{d_L} = -T_3^{e_L} = 1/2,
\\
&&
T_3^{u_R} = T_3^{d_R} = T_3^{e_R} = 0,
\\
&&
Q_{u_L} = Q_{u_R} = 2/3,
\\
&&
Q_{d_L} = Q_{d_R} = -1/3,
\\
&&
Q_{e_L} = Q_{e_R} = -1,
\\
&&
Q_{\nu_L} = 0.
\eeq
Also, $N_c = 3$, and
\beq
n_Q = n_u = n_d = n_L = n_e = 3
\eeq
are the the numbers of flavors of 
two-component quarks and leptons of each 
gauge transformation type, $(u_L, d_L)$ and $u_R$ and $d_R$ and $(\nu_L, e_L)$ and
$e_R$, respectively. The quantities $N_c$, $n_Q$, $n_u$, $n_d$, $n_L$  and $n_e$ 
are kept general in the following as a way of tagging different fermion contributions, 
although they are all equal to 3 in the Standard Model.

The 1-loop contribution is then:
\beq
\Delta^{(1)}_Z &=&
N_c (a_{u_L}^2 + a_{u_R}^2) f_1(t) + N_c 2 a_{u_L} a_{u_R} f_2(t)
+ \Bigl [
N_c (n_Q - 1)  a_{u_L}^2 
\nonumber \\ &&
+ N_c (n_u - 1)  a_{u_R}^2 +
N_c n_Q  a_{d_L}^2 + N_c n_d  a_{d_R}^2 +
n_L  (a_{e_L}^2 + a_{\nu_L}^2) + n_e  a_{e_R}^2 \Bigr ] f_1(0)
\nonumber \\ &&
+ g^2 \biggl \{ 
(4W-Z) \Bigl (\frac{W}{Z} + \frac{5}{3} + \frac{Z}{12W} \Bigr ) B(W,W) 
+ \Bigl ( \frac{4W}{Z} - \frac{4}{3} - \frac{Z}{6W} \Bigr ) A(W) 
\nonumber \\ &&
+\Bigl (\frac{4hZ - 12 Z^2 - h^2}{12W} \Bigr ) B(h,Z) 
+ \Bigl ( \frac{h-2Z}{12 W} \Bigr ) A(Z) 
+ \Bigl ( \frac{3Z-h}{12 W} \Bigr ) A(h) 
\nonumber \\ &&
+\frac{4 W^2}{Z} 
-\frac{4W}{3} 
+\frac{5Z}{9}
+\frac{hZ}{6W}
+\frac{Z^2}{18W}
\biggr \}
,
\label{eq:DeltaZ1loop}
\eeq
where the fermion 1-loop integral functions are:
\beq
f_1(t) &=& \frac{2}{3} (t-Z) B(t,t) - \frac{4}{3} A(t) + \frac{2}{9} Z - \frac{4}{3} t,
\\
f_1(0) &=& -\frac{2}{3} Z B(0,0) + \frac{2}{9} Z,
\\
f_2(t) &=& -2 t B(t,t).\phantom{\frac{8}{8}}
\eeq
The basis integrals $B(0,0)$, $B(t,t)$, $B(h,Z)$,
and $B(W,W)$, and other integral functions below, are always evaluated at the
external momentum invariant $s = Z$ and renormalization scale $Q$. 
The bottom-quark, tau-lepton, and other fermion masses have been neglected for simplicity, 
because even at 1-loop order 
they make a difference of less than 1 MeV in the real $Z$ pole mass.
However, they can easily be restored in the 1-loop part by following the example of 
the top-quark terms in the obvious way. 

The 2-loop QCD contribution can also be written in terms of 
the basis integral functions in a few lines:
\beq
\Delta^{(2),\rm QCD}_Z &=&
g_3^2 \biggl (\frac{N_c^2 -1}{4}\biggr )
\biggl [
(a_{u_L}^2 + a_{u_R}^2) F_1(t) + 2 a_{u_L} a_{u_R} F_2(t)
\nonumber \\ &&
+[(n_Q-1) a_{u_L}^2 + (n_u-1) a_{u_R}^2 + n_Q a_{d_L}^2 + n_d a_{d_R}^2]
F_1(0)
\biggr ],
\eeq
where:
\beq
F_1(t) &=&
\frac{8}{3} (Z-t) (2t-Z) M(t,t,t,t,0) + \frac{16}{3} (Z-t) \Tbar(0,t,t)
\nonumber \\ &&
+\frac{1}{3Z(4t-Z)} \Bigl [
(24 t^2 Z -24t^3 + 20 t Z^2 - 8 Z^3) B(t,t)^2
\nonumber \\ &&
+ (32 Z^2- 32 t Z -48 t^2 ) A(t) B(t,t)
+ (56 Z -24 t + 16 Z^2/t) A(t)^2
\nonumber \\ &&
- 4 (t-Z) (12 t^2 - 30 t Z + 7 Z^2) B(t,t)
+ (296 t Z - 48 t^2 - 104 Z^2) A(t)
\nonumber \\ &&
-24 t^3 + 220 t^2 Z - 141 t Z^2 + 23 Z^3
\Bigr ],
\\
F_1(0) &=& -\frac{8}{3} Z^2 M(0,0,0,0,0) - 4 Z B(0,0) - \frac{31}{3} Z,
\\
F_2(t) &=&
8t (2t-Z) M(t,t,t,t,0) + 16 t\, \Tbar(0,t,t)\phantom{\frac{u}{.}}
\nonumber \\ &&
+\frac{1}{Z(4t-Z)} \Bigl [
(8 t^3 + 4 t Z^2) B(t,t)^2
+ (16 t^2 + 80 t Z) A(t) B(t,t)
\nonumber \\ &&
+ (8t + 64 Z) A(t)^2
+ (16 t^3 - 200 t^2 Z + 36 t Z^2) B(t,t)
\nonumber \\ &&
+ (16 t^2 - 104 t Z) A(t)
+ 8 t^3 - 140 t^2 Z + 43 t Z^2
\Bigr ] .
\eeq

The 2-loop non-QCD contribution to the $Z$ boson pole squared mass 
has the form:
\beq
\Delta^{(2),{\rm non-QCD}}_Z &=& 
\sum_i c^{(2)}_i I_i^{(2)}
+ \sum_{j\leq k} c_{j,k}^{(1,1)} I_j^{(1)} I_k^{(1)}
+ \sum_j c^{(1)}_j I_j^{(1)} + c^{(0)} .
\label{eq:Delta2M2ZnonQCD}
\eeq
where the list of 1-loop basis integrals is
\beq
I^{(1)} &=& \bigl \{
A(h),\> A(t),\> A(W),\> A(Z),\> B(0,0),\> B(t,t),\> B(h,Z),\> B(W, W)
\bigr \} ,
\label{eq:onelooplist}
\eeq
and the list of necessary 2-loop basis integrals is:
\beq
I^{(2)} &=& \bigl \{
I(0, 0, h),\> I(0, 0, t),\> I(0, 0, W),\> I(0, 0, Z),\>
I(0, h, W),\> I(0, h, Z),\> 
\nonumber \\ &&
I(0, t, W),\> I(0, W, Z),\> I(h, h, h),\> I(h, t, t),\>
I(h, W, W),\> I(h, Z, Z),\> 
\nonumber \\ &&
I(t, t, Z),\> I(W, W, Z),\> M(0, 0, 0, 0, 0),\>
M(0, 0, 0, 0, W),\> M(0, 0, 0, 0, Z),\> 
\nonumber \\ &&
M(0, t, 0, t, W),\> M(0, W, 0, W, 0),\>
 M(0, W, 0, W, t),\> M(h, h, Z, Z, h),\> 
\nonumber \\ &&
M(h, t, Z, t, t),\> M(h, W, Z, W, W),\>
 M(h, Z, Z, h, Z),\> M(t, t, t, t, 0),\> 
\nonumber \\ &&
M(t, t, t, t, h),\> M(t, t, t, t, Z),\>
 M(t, W, t, W, 0),\> M(W, W, W, W, 0),\> 
\nonumber \\ &&
M(W, W, W, W, h),\> M(W, W, W, W, Z),\>
S(0, 0, h),\> S(0, 0, W),\> S(0, t, W),\> 
\nonumber \\ &&
S(h, h, Z),\> S(h, t, t),\> S(h, W, W),\>
 S(t, t, Z),\> S(W, W, Z),\> S(Z, Z, Z),\> 
\nonumber \\ &&
T(h, 0, 0),\> T(h, h, Z),\> T(h, t, t),\>
 T(h, W, W),\> T(t, 0, W),\> T(t, h, t),\> 
\nonumber \\ &&
 T(t, t, Z),\> T(W, 0, 0),\> T(W, 0, t),\>
 T(W, h, W),\> T(W, W, Z),\> T(Z, 0, 0),\> 
\nonumber \\ &&
\Tbar(0, t, t),\> \Tbar(0, W, W),\>
 U(h, Z, 0, 0),\> U(h, Z, h, Z),\> U(h, Z, t, t),\> 
\nonumber \\ &&
U(h, Z, W, W),\> U(t, t, 0, W),\>
 U(t, t, h, t),\> U(t, t, t, Z),\> U(W, W, 0, 0),\>
\nonumber \\ &&
U(W, W, 0, t),\> U(W, W, h, W),\>
U(W, W, W, Z),\> U(Z, h, h, h),\> 
\nonumber \\ &&
U(Z, h, t, t),\> 
U(Z, h, W, W),\> U(Z, h, Z, Z)
\bigr \} .
\eeq
The coefficients $c^{(2)}_i$ and $c_{j,k}^{(1,1)}$ and $c^{(1)}_j$ and 
$c^{(0)}$ are quite lengthy, so they will not be listed in print here. 
Instead, they are listed in electronic form in an ancillary file 
provided with the arXiv source for this 
article, called {\tt coefficients.txt}. 
They are ratios of polynomials of $Z$, 
$W$, $t$, $h$, and $v$. As usual, these coefficients are not unique, 
because of special identities that relate different basis integrals 
in cases where the masses are not generic.

For each of the five-propagator $M$ integrals for which analytical 
results are not available, the main TSIL Runge-Kutta evaluation function 
\verb|TSIL_Evaluate| simultaneously computes all of the subordinate 
integrals $S$, $T$, $U$ obtained by removing one or more propagator 
lines. Therefore, only 11 calls of \verb|TSIL_Evaluate| are required (in 
addition to the relatively fast evaluation of the integrals that are 
known in terms of polylogarithms), and in total the numerical 
computation takes well under 1 second on modern computer hardware.

I performed a number of stringent analytical checks on the calculation, 
similar to those described for the calculations of 
the Higgs and $W$ boson pole masses in \cite{Martin:2014cxa,Martin:2015lxa}.
First, $s^Z_{\rm pole}$ is free of poles in $\epsilon$. 
The cancellation of these poles relies on agreement between the 
divergent parts of the loop integrals performed here and the counterterm
coefficients which can be obtained from the 2-loop scalar anomalous dimension
and $\beta$ functions from refs.~\cite{MVI,MVII,Jack:1984vj,MVIII}.
Second, poles and logs of the 
Goldstone boson squared mass $G = m^2 + \lambda^2 v^2$ were checked to cancel
after the resummation described in \cite{Martin:2014bca,Elias-Miro:2014pca}.
Third, I checked the cancellations between contributions from  
unphysical vector propagator components with poles at 0 squared mass
and the corresponding Landau gauge Goldstone boson propagators. This ensures
the absence of unphysical imaginary parts of the complex pole squared mass.
Note that $\Gamma_Z = 0$ in the case $N_c = n_L = n_e = 0$. 
Next, I checked the absence of singularities in various formal limits
(none of which are close to being realized in the actual parameters of the 
Standard Model), in which one or more of the following quantities
vanish: $Z$, $W$, $t$, $h$, $t-W$, $4 t - Z$, $4W - h$, and $4 Z - h$.  
This is despite the fact that many of
the individual 2-loop coefficients
do have singularities in one or more of those cases; 
non-trivial relations between 
basis integrals are responsible for the smooth limits of the total.
Finally, I checked analytically that the complex pole squared mass is
renormalization group scale-invariant 
up to and including all terms of 2-loop order, using
\beq
Q\frac{d}{dQ} s^Z_{\rm pole} &=& 
\left [ Q \frac{\partial}{\partial Q}
- \gamma v \frac{\partial}{\partial v}
+ \sum_X \beta_{X} \frac{\partial}{\partial X} 
\right ] s^Z_{\rm pole} \>=\> 0,
\label{eq:RGinvar}
\eeq
where $\gamma$ is the Higgs anomalous
dimension, and $X = \{g, g', g_3, y_t, \lambda\}$.
In the conventions used here,
the derivatives of the 1-loop basis integrals
with respect to squared masses are listed in
eqs.~(A.5) and (A.6) of ref.~\cite{Martin:2014cxa}, while
the derivatives of the 1-loop and 2-loop
basis integrals with respect to the renormalization scale $Q$ can be found 
in eqs.~(4.7)-(4.13) of ref.~\cite{Martin:2003qz}. 
The beta functions and scalar anomalous dimension are listed
in refs.~\cite{MVI,MVII,Jack:1984vj,MVIII,FJJ}.
In the next section, a numerical check of the $Q$ invariance will be shown.

In refs.~\cite{Jegerlehner:2001fb,Jegerlehner:2002em}, a calculation of the
$Z$ boson pole mass in the pure $\MSbar$ scheme has already been given. 
However, unlike the present paper, they expanded around the tree-level definition of the 
VEV, as discussed in the Introduction above. This means that even at 1-loop order,
the results take different forms. 
The expression for the 1-loop pole squared 
mass contribution $\Delta^{(1)}_Z/16 \pi^2$ given in 
eq.~(\ref{eq:DeltaZ1loop}) above appears to
differ from the result of eq.~(B.4) of ref.~\cite{Jegerlehner:2001fb} and 
eq.~(B.3) of ref.~\cite{Jegerlehner:2002em}
by an amount 
\beq
\frac{Z}{16 \pi^2 v^2 h} \left [
-8 N_c t A(t) + 3 h A(h) + 12 W A(W) + 8 W^2 + 6 Z A(Z) + 4 Z^2
\right ] ,
\label{eq:compJKV}
\eeq
in the notation of the present paper.
There is of course no contradiction; this merely reflects the difference
between the tree-level contributions, which are $(g^2 + g^{\prime 2}) v^2/4$ in this paper
and $(g^2 + g^{\prime 2}) v^2_{\rm tree}/4$ in 
refs.~\cite{Jegerlehner:2001fb,Jegerlehner:2002em}. 
Note in particular the presence of $1/h \propto 1/\lambda$ in eq.~(\ref{eq:compJKV});
at loop order $\ell$, the use of the tree-level VEV results in terms proportional to
$1/\lambda^\ell$.
In contrast, there are no $\lambda \rightarrow 0$ singularities in the present paper.
A detailed comparison would be much more difficult at 2-loop order, 
as refs.~\cite{Jegerlehner:2001fb,Jegerlehner:2002em}
also relied on doing high-order 
expansions in $Z/h$ and $Z/t$ and $1/4 - \sin^2\theta_W$.

\section{Numerical results\label{sec:num}}
\setcounter{equation}{0}
\setcounter{figure}{0}
\setcounter{table}{0}
\setcounter{footnote}{1}

Consider a benchmark set of Standard Model $\MSbar$
parameters defined at the input renormalization scale
$Q = M_t = 173.34$ GeV:
\beq
g(M_t) &=& 0.647550,
  \label{eq:inputg}
\\
g'(M_t) &=& 0.358521,
  \label{eq:inputgp}
\\
y_t(M_t) &=& 0.93690,
  \label{eq:inputyt}
\\
g_3(M_t) &=& 1.1666,
  \label{eq:inputg3}
\\
v(M_t) &=& \mbox{246.647 GeV},
  \label{eq:inputvev}
\\
\lambda(M_t) &=& 0.12597,
  \label{eq:inputlambda}  
\eeq
The gauge couplings $g$ and $g'$ are taken to agree with ref.~\cite{Degrassi:2014sxa},
while $y_t$ and $g_3$ 
are from eqs.~(57) and (60) of version 4 of ref.~\cite{Buttazzo:2013uya}. 
The VEV $v(M_t)$ and the Higgs self-coupling were then chosen so that $M_Z$ 
agrees with the central value of eq.~(\ref{eq:MZpoleLEP}), when computed at $Q =M_Z$,
and $M_h$ agrees with the current experimental central value \cite{Aad:2015zhl} of 
$M_h = 125.09$ GeV, when computed at $Q = 160$ GeV
using the program SMH \cite{SMHwebpages} as described in ref.~\cite{Martin:2014cxa}.
With this set of input parameters, one also obtains $m^2(M_t) = -(92.890\>\,{\rm GeV})^2$ from minimization of
the Higgs potential using SMH at $Q = M_t$. 
In this way, the experimental measurements of $M_Z$ and $M_h$ can be used to 
obtain the Higgs potential parameters. The choice of $Q=160$ GeV for computing
$M_h$ was explained in ref.~\cite{Martin:2014cxa}; at this scale the
effects of top-quark loops in the neglected electroweak 3-loop parts
should be not too large. The lower choice of $Q = M_Z$
for computing $M_Z$ is somewhat arbitrary.
One also obtains a $W$ boson pole mass of $M_W = 80.329$ GeV, when computed at $Q = M_W$, 
using the calculation described in \cite{Martin:2015lxa}. 
This translates into a Breit-Wigner mass of $M_W^{\rm exp} = 80.356$ GeV, 
using the analog of eq.~(\ref{eq:poleBW}) above. (Somewhat coincidentally, 
this agrees with the value found in ref.~\cite{Degrassi:2014sxa} to within 1 MeV,
although that calculation uses a different scheme.)

\begin{figure}[!p]
\includegraphics[width=0.6\linewidth,angle=0]{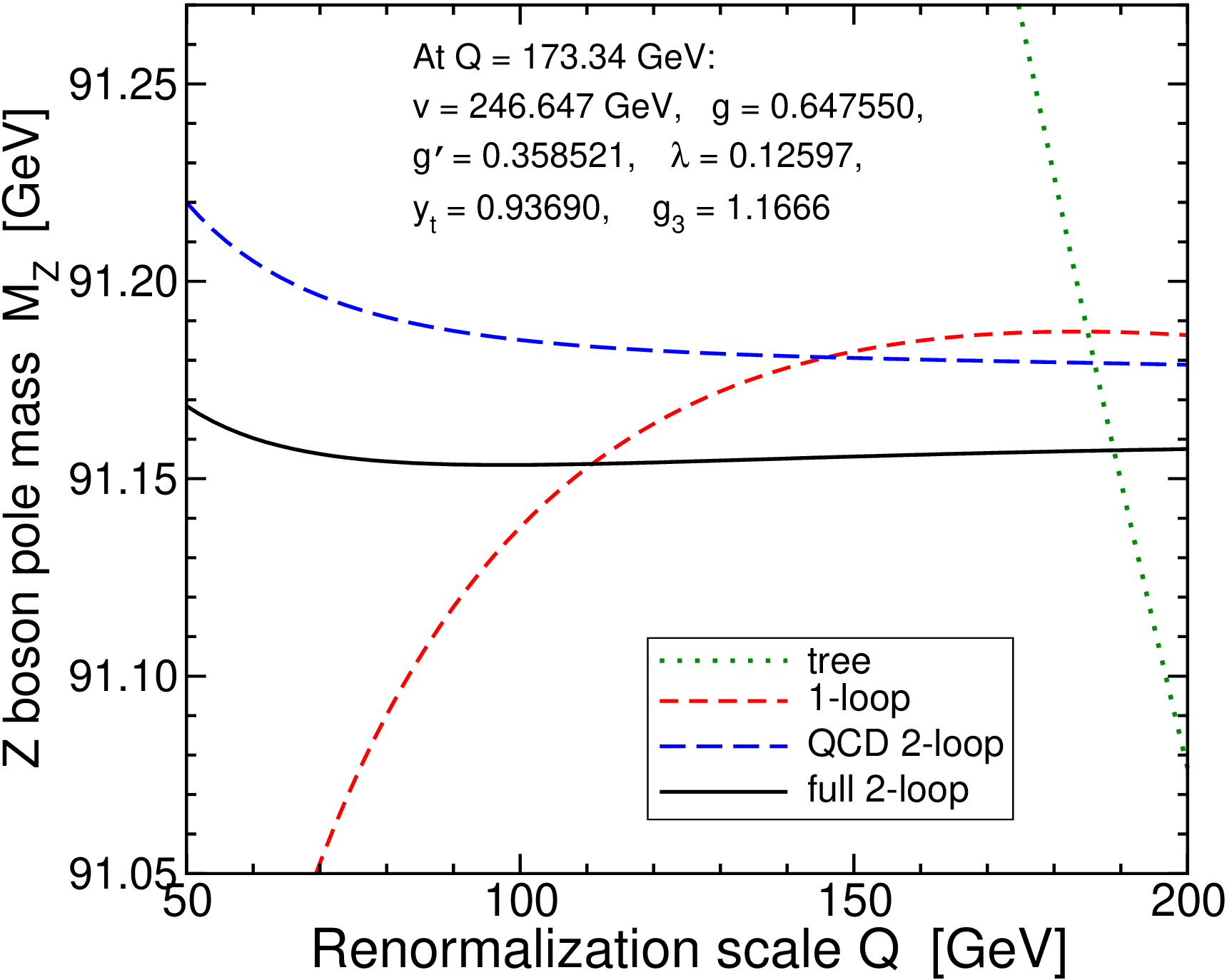}
\begin{minipage}[]{0.95\linewidth}
\caption{The computed pole mass $M_Z$ of the $Z$ boson, defined by
$s^Z_{\rm pole} = M_Z^2 - i \Gamma_Z M_Z$,
as a function of the renormalization scale $Q$ at which
it is computed, in various approximations. 
The dotted (green) line is the tree-level result $Z$, 
the short-dashed (red) line is the 1-loop result, the long-dashed (blue) line
is the result from the 1-loop and 2-loop QCD contribution, while the
solid (black) line is the full 2-loop order result. 
The input parameters $g, g', y_t, g_3, \lambda$, and $v$ 
at the renormalization 
scale $Q$ are obtained by running 3-loop renormalization group running, 
starting from eqs.~(\ref{eq:inputg})-(\ref{eq:inputlambda}).
Note that the usual Breit-Wigner mass $M_Z^{\rm exp}$ is 0.0341 GeV larger
than the $M_Z$ shown here.
\label{fig:MZQ}}
\end{minipage}
\end{figure}

\begin{figure}[!p]
\includegraphics[width=0.6\linewidth,angle=0]{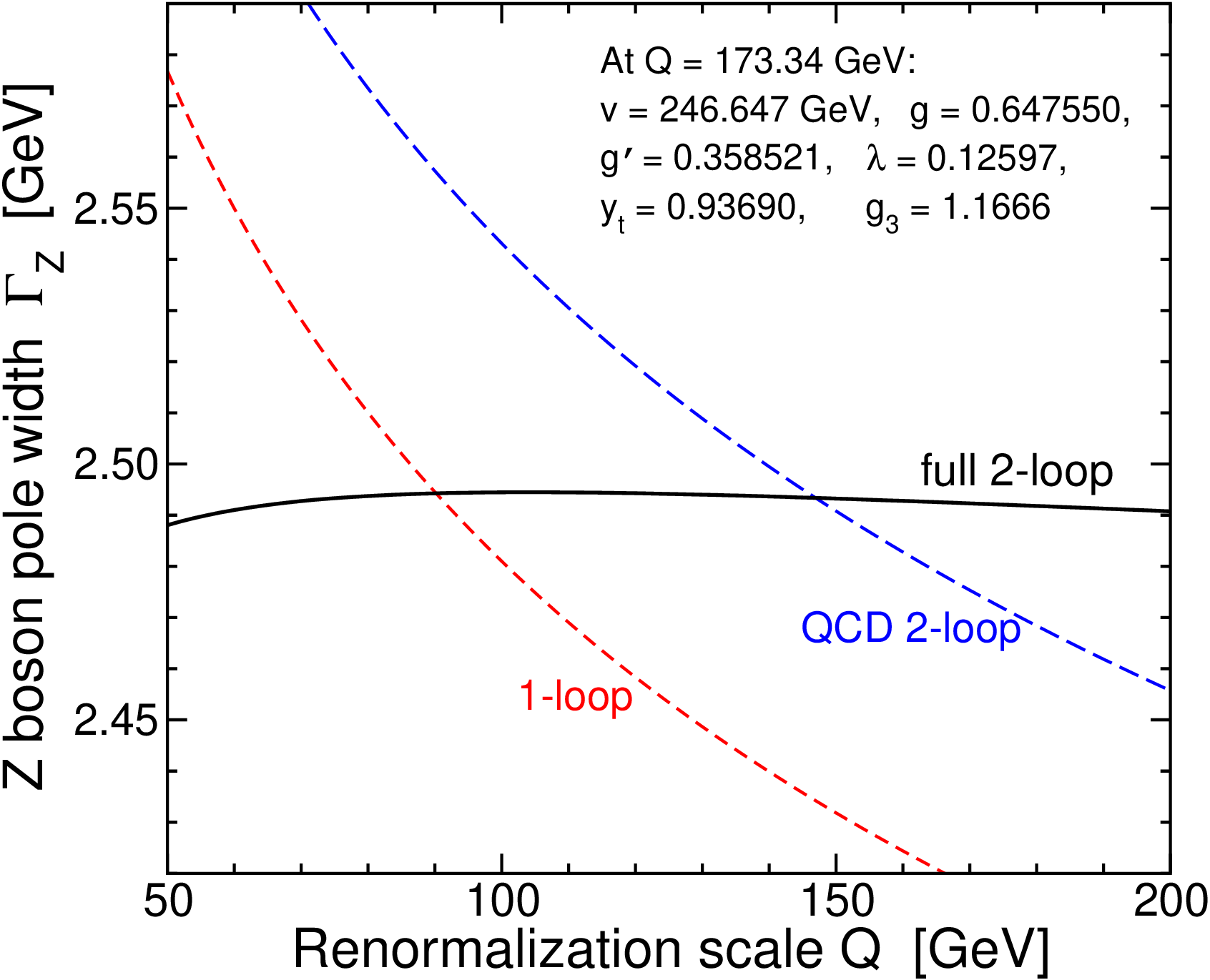}
\begin{minipage}[]{0.95\linewidth}
\caption{The computed width $\Gamma_Z$ of the $Z$ boson, defined in terms of the
complex pole squared mass $s^Z_{\rm pole} = M_Z^2 - i \Gamma_Z M_Z $,
as in Figure \ref{fig:MZQ}.
The short-dashed (red) line is the 1-loop result, 
the long-dashed (blue) line
is the result from the 1-loop and 2-loop QCD contribution, and the 
solid (black) line is the full 2-loop order result. 
\label{fig:GammaZQ}}
\end{minipage}
\end{figure}

The dependences of the computed pole mass parameters
$M_Z$ and $\Gamma_Z$ on the choice of $Q$ are shown in figures \ref{fig:MZQ}
and \ref{fig:GammaZQ}, in various approximations.
These graphs are made by running the input parameters 
$g$, $g'$, $y_t$, $g_3$, $\lambda$, and $v$, using their 3-loop beta 
functions \cite{Chetyrkin:2013wya,Bednyakov:2013eba},
from the input scale $M_t$ to the scale $Q$ on the horizontal axis, 
where $s^Z_{\rm pole}$ is computed. 
In an idealized case that $s^Z_{\rm pole}$ is computed to sufficiently 
high order in perturbation theory, $M_Z$ and $\Gamma_Z$ would be independent of $Q$. 
Therefore the $Q$-independence is a check on the calculation.
I find that the calculated 2-loop value of $M_Z$ varies by only about $\pm 2$ MeV
from its median value, over the range 70 GeV $<Q<$ 200 GeV. 
Below $Q = 70$ GeV, the scale dependence is much stronger.
The scale dependence is smallest for $Q$ near 100 GeV, where the computed
$M_Z$ has its minimum, but this does 
not necessarily mean that this is the best renormalization scale; 
only a higher-order calculation can reduce the theoretical uncertainty.

With regard to the width $\Gamma_Z$, the scale dependence of the full 2-loop result
is again about 
$\pm 2$ MeV from the median value over the same range 70 GeV $<Q<$ 200 GeV.
Note that here, including only the QCD part of the 2-loop contribution does not
actually reduce the scale dependence much compared to the 1-loop result. This is
because most of the $Q$ dependence in the width arises from the runnings of the
VEV and the 
electroweak couplings of the $Z$ boson to the fermions into which it decays, 
and these are independent of QCD at the leading (1-loop) order.
The result for $\Gamma_Z$ is consistent with, and slightly lower than the central value of, the experimental range \cite{RPP}
$\Gamma_Z = 2.4952 \pm 0.0023$ GeV.
Of course, there are much better ways to calculate $\Gamma_Z$, 
because the imaginary part of the 2-loop complex pole mass 
really corresponds to only a 1-loop calculation of the $Z$ width.
(Moreover, the inclusion of bottom-quark mass effects, neglected
above for simplicity, has a larger effect on $\Gamma_Z$ than on 
$M_Z$, and will decrease the former by an amount of order 2 MeV due to kinematics. 
There is a significant uncertainty in estimating this reduction in the
imaginary part of the $Z$ complex pole mass, because
of the large difference between the pole 
and running bottom quark masses.)

\begin{figure}[!t]
\includegraphics[width=0.6\linewidth,angle=0]{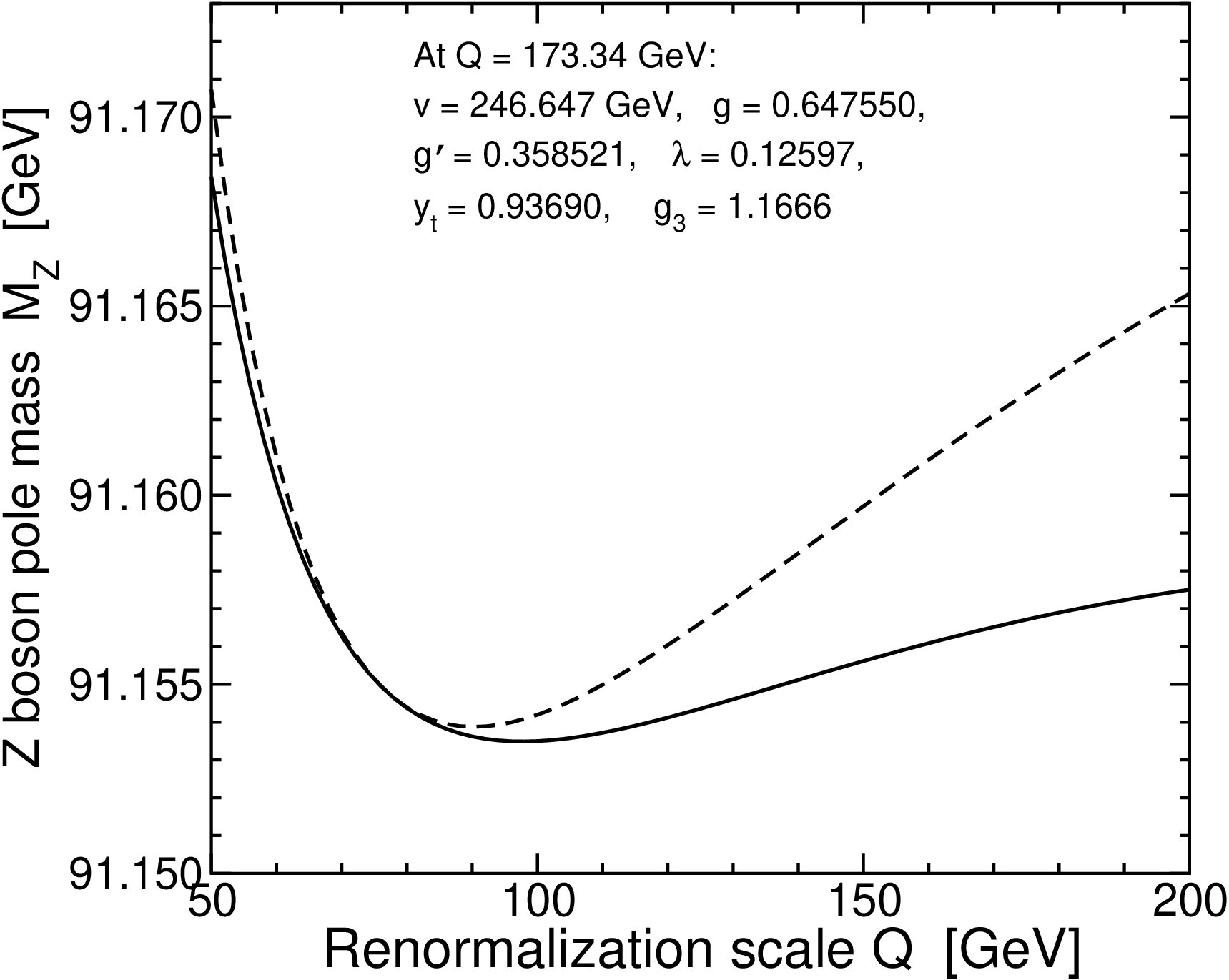}
\begin{minipage}[]{0.95\linewidth}
\caption{A close-up of the renormalization scale dependence of
the computed pole mass $M_Z$. 
The solid line is the 2-loop result, just as in Figure \ref{fig:MZQ}.
The input parameters were chosen so that at $Q=M_Z$ the computed pole mass agrees
with the experimental central value $M_Z = 91.1535$ GeV 
from eq.~(\ref{eq:MZpoleLEP}).
The dashed line is the same, but after expanding the 1-loop part in
the $\MSbar$ squared mass $t$ to linear order 
about the value $T = (173.34\>\,{\rm GeV})^2$, 
using eqs.~(\ref{eq:f1tZalt}) and (\ref{eq:f2tZalt}). 
This provides an alternate
consistent 2-loop order result. The two approximations 
agree near the scale $Q = 77$ GeV
where $t=T$ (the top-quark running and pole masses coincide).
Note that the Breit-Wigner mass $M_Z^{\rm exp}$ is 0.0341 GeV larger
than the pole mass $M_Z$ shown here.
\label{fig:MZQalt}}
\end{minipage}
\end{figure}

It is important to keep in mind that the renormalization scale dependence
only provides a lower bound on the theoretical error. 
Another way of investigating the robustness of the calculation is 
to take the running top-quark squared mass $t$ in the 1-loop part 
eq.~(\ref{eq:DeltaZ1loop}) and perform an expansion around an arbitrary 
value $T$ that can be considered to differ from $t$ by an amount that is 
parametrically of 1-loop order. An obvious choice is to take $T$ to be 
the (real part of the) top-quark pole squared mass. It makes sense to do 
this in particular for the 1-loop contribution, because the top quark 
mass appears only in propagators at this order, not as a vertex Yukawa 
coupling. Expanding, one finds:
\beq
f_1(t) &=& f_1(T) + (t-T) \left [
(4 T - 2 Z) B(T,T) - 4 A(T) - 12 T + 4 Z \right ]/(4T-Z) 
+ \ldots , \phantom{xx}
\label{eq:f1tZalt}
\\
f_2(t) &=& f_2(T) + (t-T) \left [
(2Z - 12 T) B(T,T) - 4 A(T) + 4 T \right ]/(4T-Z) 
+ \ldots .\phantom{xx}
 \label{eq:f2tZalt}
\eeq
I have checked that if these expansions were 
continued to include order $(t-T)^3$, 
then the results for the $Z$ pole squared mass 
would be nearly indistinguishable from the original result obtained directly
from the un-expanded $f_1(t)$ and $f_2(t)$. 
However, by instead 
keeping the expansion only at first order in
$(t-T)$ as shown, one obtains an alternative consistent 2-loop order result 
for the $Z$ pole squared mass, since $t-T$ is
to be treated as formally of 1-loop order. This alternative consistent 2-loop order
result is numerically different, with the difference giving an indication 
of the magnitude of the error made in
terminating perturbation theory at 2-loop order. 
The result of using eqs.~(\ref{eq:f1tZalt}) and (\ref{eq:f2tZalt})
compared to the original un-expanded $f_1(t)$ and $f_2(t)$ 
is shown in Figure \ref{fig:MZQalt}.
We see that the alternate consistent 2-loop result, shown as the dashed line,
has a significantly worse scale dependence, especially at larger $Q$. This
suggests that the scale dependence of $M_Z$ found in the original calculation
(the solid line) is 
actually accidentally small, and probably 
underestimates the theoretical error. A very similar behavior was found for the $W$ boson
mass in ref.~\cite{Martin:2015lxa}.

\section{Outlook\label{sec:outlook}}
\setcounter{equation}{0}
\setcounter{figure}{0}
\setcounter{table}{0}
\setcounter{footnote}{1}

In this paper I have provided a full 2-loop calculation of the $Z$ boson complex pole
square mass in the pure $\MSbar$ scheme, to go along with similar results for the
$W$ boson \cite{Martin:2015lxa} and the 
Higgs boson \cite{Martin:2014cxa} using the same renormalization 
scheme and the same
definition of the VEV. These calculations are an alternative to the 
on-shell scheme results 
that have been widely used for precision studies in the Standard Model, in which $M_Z$
instead plays the role of an input parameter. 

The ultimate goal should be to obtain results in which the theoretical error is very small
compared to present and projected 
experimental errors. 
The previous section shows that this is certainly not obtained using 
just the full 2-loop calculation, as the scale dependence is comparable to
the experimental errors, and the theoretical error is probably somewhat larger. 
There is no compelling evidence or argument that the subset of 3-loop
contributions that are QCD and top-Yukawa enhanced will be enough to ensure 
the dominance of
experimental errors over theoretical errors. At 2-loop order, one can see from the
benchmark example of Figure \ref{fig:MZQ} that the QCD contribution has a 
much larger scale dependence, but not a much larger magnitude, than the non-QCD 
contributions, except
for smaller choices of the renormalization scale $Q$ where the top-enhanced QCD 
corrections are big. 
The same thing was noted in the comparable results
for the $W$ boson in \cite{Martin:2015lxa}. It is therefore reasonable to conclude that
complete 3-loop calculations will be necessary, 
providing a worthy challenge for future work.

\vspace{0.2cm}

\noindent {\it Acknowledgments:} 
This work was supported in part by the National
Science Foundation grant number PHY-1417028. 


\end{document}